\documentclass{ptephy_v1}

\preprintnumber{OU-HET-1071} 

\usepackage{ascmac}

\newtheorem{theorem}{Theorem}

\begin{document}

\title{Building bulk from Wilson loops}


\author{Koji Hashimoto}
\affil{Department of Physics, Osaka University\\
Toyonaka, Osaka 560-0043, JAPAN
\email{koji@phys.sci.osaka-u.ac.jp}}

%
%
%

\begin{abstract}%
We provide formulas for holographically building a bulk metric from given expectation values of
rectangular Wilson loops. As a corollary, we prove that any confining quark potential necessarily leads to
the existence of a bulk IR bottom.
\end{abstract}


\maketitle

\setcounter{tocdepth}{1}
\tableofcontents

\section{Introduction: building a bulk from QFT data}

Building a bulk gravity spacetime from a given quantum field theory (QFT) data is one of the central issues in the holographic principle and
the AdS/CFT correspondence \cite{Maldacena:1997re,Gubser:1998bc,Witten:1998qj}.
The difficulty resides in its inverse problem nature. 
Normally in AdS/CFT, once a bulk spacetime or a bulk gravity system is given, 
then boundary QFT data such as correlators are calculated by solving classical physics in the bulk.
On the other hand, building a bulk from a given QFT data is going backward,
and going from lower dimensions to higher dimensions\footnote{
The easiest analogue of this difficulty is the problem of finding a quantum mechanical potential when 
only energy eigenvalues are provided. QFTs are infinite-dimensional and many-body generalization of the quantum mechanics,
which suggests the difficulty.}, by decoding the holographic principle.
The issue of building a bulk is concerned with the unrevealed mechanism of the emergent holographic dimension,
which also signals the importance and the difficulty of the issue.

There have been a great advance in research on building a bulk spacetime metric from given boundary QFT data. 
The holographic renormalization \cite{deHaro:2000vlm} provides a bulk metric for a given expectation value of a QFT 
energy momentum tensor. A bulk reconstruction using bulk geodesics and light cones 
\cite{Hammersley:2006cp,Hubeny:2006yu,Engelhardt:2016wgb,Engelhardt:2016crc,Roy:2018ehv,Burda:2018rpb,Hernandez-Cuenca:2020ppu}
uses divergent behavior of QFT correlators.
A bulk-building method using holographic entanglement entropy \cite{Ryu:2006bv,Ryu:2006ef}
 has been developed extensively \cite{Hammersley:2007ab,Bilson:2008ab,Bilson:2010ff,Hubeny:2012ry,Balasubramanian:2013rqa,Balasubramanian:2013lsa,Myers:2014jia,Czech:2014wka,Czech:2014ppa,Gentle:2015cfp,Freedman:2016zud,Saha:2018jjb,Bao:2019bib,Cao:2020uvb,Jokela:2020auu,Agon:2020mvu}. Relatedly, the identification of the bulk with tensor networks \cite{Swingle:2009bg,Pastawski:2015qua} through the entanglement properties
has been utilized to build a bulk space
\cite{Swingle:2012wq,Bao:2018pvs,Milsted:2018san,Bao:2019fpq}.

In spite of these developments, 
there are obstacles to use the existing technologies to build a bulk spacetime metric from genetic QFT's. 
This can be manifested in a question: what is a gravity dual of the Yang-Mills theory? 
Generically in AdS/CFT, the boundary QFT is strongly coupled, so the preparation of any QFTdata needs
non-perturbative methods such as lattice simulations. Unfortunately, the real-time correlators and the entanglement entropies
used above are currently difficult to evaluate on the lattice. While energy momentum tensors were evaluated on the lattice recently,
the holographic renormalization assumes bulk Einstein equations. 

In this paper, we use Wilson loops as the QFT data to build a bulk spacetime metric. 
The Wilson loops in lattice simulations have been well studied and are well defined.
The gravity dual of the Wilson loops \cite{Maldacena:1998im,Rey:1998ik,Brandhuber:1998bs,Rey:1998bq} is
known and does not assume the bulk Einstein equations.
And fortunately, it is a minimal surface in the bulk, thus the bulk-building technologies 
developed for the entanglement entropy \cite{Bilson:2008ab,Bilson:2010ff} 
with the Ryu-Takayanagi surface can be employed.\footnote{
For the interplay between the Ryu-Takayanagi surface and the holographic Wilson loop in thermalization
and confinement, see
\cite{Balasubramanian:2010ce, Balasubramanian:2011ur,Kol:2014nqa,Bao:2019hwq}.}

Our target bulk spacetime is assumed to be given by 
a bulk metric in string frame in any spacetime dimensions ($\geq 3$),
\begin{align}
ds^2 = -f(\eta) dt^2 + g(\eta) d\vec{x}^2 + d\eta^2 \, ,
\label{m}
\end{align}
where $\eta$ is the emergent holographic radial coordinate. 
Generically
the bulk metric can always be cast into this form, when the QFT is invariant under spacetime translation 
and space rotation, which is often favored in measuring Wilson loops on the lattice in the continuum limit.
We take the gauge $g_{\eta\eta}=1$
so that the proper distance along the radial direction is
measured simply by the $\eta$ coordinate.
We also assume that $f(\eta)$ and $g(\eta)$ are monotonic functions of $\eta$, 
and $f(\infty)=g(\infty)=\infty$ which defines the boundary of the bulk geometry.

According to \cite{Maldacena:1998im,Rey:1998ik,Brandhuber:1998bs,Rey:1998bq}, 
once the bulk metric \eqref{m} is given, the quark potential $E$
and the interquark distance $R$ are calculated by a Nambu-Goto string hanging down in the bulk from the boundary, 
as
\begin{align}
E(\eta_0) & = \frac{1}{\pi \alpha'}\int_{\eta_0}^\infty d\eta \, 
\sqrt{f(\eta)} \sqrt{\frac{f(\eta)g(\eta)}{f(\eta)g(\eta)-f(\eta_0)g(\eta_0)}} \, , 
\label{E}
\\
R(\eta_0) & = 2\int_{\eta_0}^\infty d\eta \, 
\frac{1}{\sqrt{g(\eta)}} \sqrt{\frac{f(\eta_0)g(\eta_0)}{f(\eta)g(\eta)-f(\eta_0)g(\eta_0)}} \, ,
\label{R}
\end{align}
with a parameter $\eta_0$. 
This $\eta=\eta_0$ is the deepest bulk location of the smooth tip of the string. 
Eliminating $\eta_0$ from \eqref{E} and \eqref{R} provides the quark potential $E(R)$
in the boundary QFT.

The quark potential $E(R)$ is related to the expectation value $\exp[-E(R)T]$ of a temporal Wilson loop
of a rectangular path of the size $T \times R$, in the infinite limit of the time domain $T \to \infty$. 
In the same manner, another observable available is a spatial Wilson loop of the size $R_s \times R'$,
which defines $E_s(R_s)$ in the similar limit $R' \to \infty$. The formula for $E_s(R_s)$ is
the same as \eqref{E} and \eqref{R} except for replacing $f$ by $g$.

Our goal in this paper is to provide an inverse formula: given $E(R)$ and $E_s(R)$,  
build $f(\eta)$ and $g(\eta)$. In particular, at zero temperature, since the Lorentz symmetry shows  
$f=g$, we need only $E(R)$ to build a bulk metric. Furthermore, our formula derives a theorem
stating that any confining quark potential results in the existence of an IR bottom
of the built bulk geometry.

The organization of this paper is as follows.
In Section \ref{sec:2}, we provide our analytic formulas for building a bulk spacetime from a given Wilson loop data.
In Section \ref{sec:3}, we use the formulas with a power-law quark potential, to find that the built bulk is 
near-horizon geometries of D$p$-branes.
In Section \ref{sec:4}, we provide a theorem that the confinement leads to a bulk IR bottom. 
Section \ref{sec:5} is for discussions on various aspects of our formulas.

\section{Formulas}
\label{sec:2}

The formulas we obtain to build a bulk metric \eqref{m} at zero temperature and at non-zero temperature
are as follows. 

\vspace{3mm}

\begin{itembox}[l]{Zero temperature case}

Given a quark potential $E(R)$, solve
\begin{align}
f_0 = 2\pi \alpha' \frac{dE(R)}{dR}
\label{f0}
\end{align}
to get $R$ as a function of $f_0$. 
Then substitute it to the following differential equation
\begin{align}
\frac{d\eta(f)}{df}
= \frac{1}{\pi}\sqrt{f}\frac{d}{df}
\int_\infty^f df_0 \frac{R(f_0)}{\sqrt{f_0^2-f^2}} \, .
\label{deta}
\end{align}
Integrate this to find $\eta=\eta(f)$.
Finally, invert it to find a bulk metric $f(\eta)$.

\end{itembox}
\vspace{3mm}

\begin{itembox}[l]{Non-zero temperature case}

Given a potential $E_s(R_s)$ for a spatial Wilson loop and $E(R)$ for a temporal Wilson loop, solve
\begin{align}
g_0 = 2\pi \alpha' \frac{dE_s(R_s)}{dR_s}\, , 
\quad
h_0 = 2\pi \alpha' \frac{dE(R)}{dR}\, , 
\label{gh}
\end{align}
to get $R_s(g_0)$ and $R(h_0)$. 
First, substitute $R_s(g_0)$ to the differential equation
\begin{align}
\frac{d\eta(g)}{dg}
= \frac{1}{\pi}\sqrt{g}\frac{d}{dg}
\int_\infty^g dg_0 \frac{R_s(g_0)}{\sqrt{g_0^2-g^2}} \, .
\label{g}
\end{align}
Integrate it to find $\eta=\eta(g)$.
Invert it to find a bulk metric component $g(\eta)$. Then substitute the explicit $g(\eta)$ and also $R(h_0)$ to
the differential equation
\begin{align}
\frac{d\eta(h)}{dh}
= \frac{1}{\pi}\sqrt{g(\eta(h))}\frac{d}{dh}
\int_\infty^h dh_0 \frac{R(h_0)}{\sqrt{h_0^2-h^2}} \, .
\label{h}
\end{align}
Solve this to find $\eta(h)$, which is inverted to $h(\eta)$. Then obtain
another component of the bulk metric as 
$f(\eta) = h(\eta)^2/g(\eta)$.

\end{itembox}


\begin{proof}
First, let us provide a proof of our formulas \eqref{f0} and \eqref{deta} for the case of zero temperature,
$f(\eta)=g(\eta)$. 
We follow the strategy developed in \cite{Bilson:2008ab,Bilson:2010ff} for the entanglement entropy.
To simplify equations, we rewrite \eqref{E} and \eqref{R} with a new notation, 
\begin{align}
\tilde{E}\equiv \pi \alpha' E\, , 
\quad
\tilde{R} \equiv R/2 \, , \quad
V(\eta) \equiv f(\eta)^2 \, , \quad
f_0 \equiv f(\eta_0) \, .
\end{align}
 Then \eqref{E} and \eqref{R} are written as
\begin{align}
\tilde{E} = \int_{\eta_0}^\infty
d\eta \frac{V^{3/4}}{\sqrt{V-f_0^2}} 
\,  ,
\quad
\tilde{R} = \int_{\eta_0}^\infty
d\eta \frac{V^{-1/4}f_0}{\sqrt{V-f_0^2}} \, .
\label{etrt}
\end{align}
We shall use the following equalities:
\begin{align}
\frac{d}{d\eta}\left(\sqrt{V-f_0^2}\right) = \frac12 \frac{V'}{\sqrt{V-f_0^2}} \, , \quad
\frac{d}{d\eta}\left(\arctan\frac{\sqrt{V-f_0^2}}{f_0}\right) = \frac12 \frac{V' f_0}{V\sqrt{V-f_0^2}} \, . 
\end{align}
Making a partial integration, we find
\begin{align}
\tilde{E} 
&=
\int_{\eta_0}^\infty
\!\!\!d\eta \,
\frac12 \frac{V'}{\sqrt{V-f_0^2}} \frac{2V^{3/4}}{V'}
\nonumber \\
&
= 
\int_{\eta_0}^\infty
\!\!\!d\eta \frac{d}{d\eta}\left(\sqrt{V-f_0^2}\right) \frac{2V^{3/4}}{V'}
\nonumber \\
&
= 
-\int_{\eta_0}^\infty \!\!\!d\eta \sqrt{V-f_0^2}
\left(\frac{2V^{3/4}}{V'}\right)'
+
\left(\sqrt{V-f_0^2}\frac{2V^{3/4}}{V'}\right)
\biggm|_{\eta=\infty}
-
\left(\sqrt{V-f_0^2}\frac{2V^{3/4}}{V'}\right)
\biggm|_{\eta=\eta_0}
\nonumber \\
&
= 
-\int_{\eta_0}^\infty \!\!\!d\eta \sqrt{V-f_0^2}
\left(\frac{2V^{3/4}}{V'}\right)'
+
\left(\sqrt{V-f_0^2}\frac{2V^{3/4}}{V'}\right)
\biggm|_{\eta=\infty}
\, .
\label{et}
\end{align}
In the last equality we used $V(\eta=\eta_0) = f(\eta_0)^2 = f_0^2$.
In the same manner, we find
\begin{align}
\tilde{R}
= 
-\int_{\eta_0}^\infty \!\!\!d\eta 
\left(
\arctan\frac{\sqrt{V-f_0^2}}{f_0}
\right)
\left(\frac{2V^{3/4}}{V'}\right)'
+
\left(
\left(
\arctan\frac{\sqrt{V-f_0^2}}{f_0}
\right)\frac{2V^{3/4}}{V'}\right)\biggm|_{\eta=\infty} \, .
\label{rt}
\end{align}
Comparing \eqref{et} with \eqref{rt}, we find
\begin{align}
\frac{d \tilde{R}}{df_0} = \frac1{f_0}
\frac{d \tilde{E}}{df_0} \, .
\end{align}
This is equivalent to 
\begin{align}
f_0 = \frac{d\tilde{E}}{d\tilde{R}}\, , 
\end{align}
which proves \eqref{f0}.

Next, we shall prove \eqref{deta}. 
We rewrite $\tilde{R}$ in \eqref{etrt} as follows:
\begin{align}
\tilde{R}(f_0) = 
\int_{f_0}^{f(\eta=\infty)}df \, \left[\frac{d\eta(f)}{df} f^{-1/2}\right] \frac{f_0}{\sqrt{f^2-f_0^2}} \, .
\label{tr2}
\end{align}
Then, using the following formula
\begin{align}
F(x) = \int_x^a \!\!\!dt \, y(t) \frac{x}{\sqrt{t^2-x^2}}
\qquad
\Rightarrow
\qquad
y(t) =\frac{-2}{\pi} \frac{d}{dt}
\int_t^a \!\!\!dx \, \frac{F(x)}{\sqrt{x^2-t^2}} \, ,
\label{formula}
\end{align}
which can be proven by just substituting the first expression $F(x)$ to the last integral, 
the equation \eqref{tr2} can be inverted and 
we find
\begin{align}
\frac{d\eta(f)}{df} f^{-1/2}
= \frac{-2}{\pi}\frac{d}{df}
\int^{f(\eta=\infty)}_f d f_0 \, 
\frac{\tilde{R}(f_0)}{\sqrt{f_0^2-f^2}} \, .
\end{align}
Using $f(\infty)=\infty$, this proves \eqref{deta}. 


Next, let us provide a proof of the formulas \eqref{gh}, \eqref{g} and \eqref{h} of the non-zero temperature case.
The proof for \eqref{g} is exactly the same as the zero temperature case above. We simply replace $f$ by $g$,
then \eqref{g} with the first equation of \eqref{gh} is proven. Next, to prove \eqref{h}, we 
define
\begin{align}
V \equiv f(\eta)g(\eta)
\end{align}
and rewrite $\tilde{E}$ and $\tilde{R}$ as
\begin{align}
\tilde{E} = \int_{\eta_0}^\infty
d\eta \frac{1}{\sqrt{g(\eta)}}\frac{V(\eta)}{\sqrt{V(\eta)-V(\eta_0)}} 
\,  ,
\quad
\tilde{R} = \int_{\eta_0}^\infty
d\eta \frac{1}{\sqrt{g(\eta)}}\frac{\sqrt{V(\eta_0)}}{\sqrt{V(\eta)-V(\eta_0)}} \, .
\label{etrt2}
\end{align}
Then, putting $V(\eta_0) \equiv h_0^2$, and using exactly the same trick of partial integration, we find 
\begin{align}
h_0 = \frac{d\tilde{E}}{d\tilde{R}}
\end{align}
which is the second equation of \eqref{gh}. Writing $h(\eta) \equiv \sqrt{V(\eta)}$, \eqref{etrt2} can be written as
\begin{align}
\tilde{R} = \int_{\eta_0}^\infty d\eta \frac{1}{\sqrt{g(\eta)}}\frac{h_0}{\sqrt{h(\eta)^2-h_0^2}} \, .
\end{align}
Changing the integration variable to $h$ and using the inversion formula \eqref{formula} gives
\begin{align}
\frac{d\eta(h)}{dh} \frac{1}{\sqrt{g(\eta(h))}}
= \frac{-2}{\pi} \frac{d}{dh}
\int^{h(\eta=\infty)}_h
dh_0 \frac{\tilde{R}(h_0)}{\sqrt{h_0^2-h^2}}\, ,
\end{align}
which is nothing but \eqref{h}. 
\end{proof}

\section{Rebuilding near horizon geometries}
\label{sec:3}

In this section, to demonstrate the applicability of our formulas \eqref{deta} and \eqref{f0}, 
we adopt, as the input Wilson loop, a simple power-law potential
\begin{align}
E(R)= -\frac{c}{2\pi\alpha'} \frac{1}{R^{n-1}}
\label{genn}
\end{align}
with\footnote{When $n<2$, the built metric component $f(\eta)$ does not satisfy $f(\eta=\infty)=\infty$
which is our assumption, as we will see below. 
However, a similar analysis is possible even without the assumption.} 
$n\geq 2$ and with a positive constant $c$. 
In the following we find that our formulas build BPS near-horizon geometries of D$p$-branes
(except for the internal space part).

Let us start with the case of $n=2$, 
\begin{align}
E(R)=-\frac{c}{2\pi\alpha'} \frac{1}{R} \, .
\label{confE}
\end{align}
This corresponds to Wilson loops of conformal field theories (CFTs)
at zero temperature, because of the scaling symmetry. We will see that the formula \eqref{deta} with \eqref{f0} builds an AdS spacetime, as expected.
First, \eqref{f0} is solved as
\begin{align}
R= \sqrt{c/f_0} \, .
\end{align}
We substitute this to the right hand side of \eqref{deta}, then changing the variable of integration 
to $x = f_0/f$ gives
\begin{align}
\frac{1}{\pi}\sqrt{f}\frac{d}{df}
\int_\infty^f df_0 \frac{R(f_0)}{\sqrt{f_0^2-f^2}} 
=\frac{1}{\pi}\sqrt{f}\frac{d}{df}
\frac{\sqrt{c}}{\sqrt{f}}
\int_\infty^1 dx \frac{1}{\sqrt{x(x^2-1)}} 
= \frac{\Gamma(5/4)}{\Gamma(3/4)\sqrt{\pi}}\sqrt{c} \frac{1}{f} \, .
\end{align}
Thus integrating the differential equation \eqref{deta} gives
\begin{align}
f(\eta)= \exp \left[
\frac{2}{L}(\eta - \mbox{const.})
\right]
\label{AdS}
\end{align}
with
\begin{align}
L = \frac{2\Gamma(5/4)\sqrt{c}}{\Gamma(3/4)\sqrt{\pi}} \, .
\label{AdSr}
\end{align}
This \eqref{AdS} is nothing but an AdS spacetime, with $L$ being its AdS radius. 
Note that this rebuilt AdS is not necessarily five-dimensional.
Starting with the conformal potential \eqref{confE} in any CFT in $d$ dimensions, 
an AdS${}_{d+1}$ spacetime is obtained in the bulk.

As a trivial check of our formulas, we look at our AdS radius \eqref{AdSr} at $d=4$. 
Using the explicitly known result 
\begin{align}
E = -\frac{2\sqrt{2}\Gamma(3/4)^2}{\Gamma(1/4)^2} \sqrt{\lambda} \frac{1}{R} \, 
\end{align}
for
the Wilson loop \cite{Maldacena:1998im} evaluated by an AdS${}_5$ through the AdS/CFT correspondence,
we calculate the right hand side of \eqref{AdSr}, to find
\begin{align}
L = (2\lambda)^{1/4} \sqrt{\alpha'} \, .
\end{align}
This is the standard AdS/CFT dictionary, relating the AdS radius and $\alpha'$ 
with the 't Hooft coupling $\lambda$ of the ${\cal N}=4$ supersymmetric Yang-Mills theory in 4 dimensions. 

Next, let us consider the general case \eqref{genn} with $n>2$. 
Substituting it to \eqref{f0}, we find 
$R \propto f_0^{-1/n}$. Then, \eqref{deta} gives
\begin{align}
\frac{d\eta(f)}{df} \propto f^{-1/n-1/2} \, .
\end{align}
When $n> 2$, integrating this finds $(\eta(f)-\mbox{const.}) \propto f^{-1/n+1/2}$, which gives
the bulk metric
\begin{align}
f(\eta) \propto (\eta-\mbox{const.})^\frac{2n}{n-2} \, .
\end{align}
So, we conclude that the power-law quark potential \eqref{genn} for $n>2$ leads to a power-law $f(\eta)$ of the bulk 
geometry, and for $n=2$ it leads to an AdS spacetime.

The power-law behavior of $f(\eta)$ is popularly known for the near horizon geometry of D$p$-branes \cite{Itzhaki:1998dd},
\begin{align}
ds^2 = \eta^\frac{2(7-p)}{p-3}\left(-dt^2+\sum_{i=1}^p (dx^i)^2\right) + d\eta^2 + \mbox{(internal space)} \, ,
\end{align}
where we ignored numerical coefficients. 
If we employ this expression, the relation between the power $n$ of the quark potential and the spatial dimensionality $p$ 
of the QFT is found as $n = (7-p)/(5-p)$.
This relation is the one obtained in the holographic Wilson loop calculations in \cite{Brandhuber:1998er}, so it serves as
another check of our formulas.

\section{Theorem: confinement leads to a bulk IR bottom }
\label{sec:4}

In \cite{Kinar:1998vq}, a beautiful theorem about the IR behavior of the metric $f(\eta)$
and confinement was provided. It states that when there exists an ``IR bottom" $\eta=0$
such that $f(\eta)$ can be expanded around it as $f(\eta)=f(0) + a \eta^k$ with $f(0)>0$ and $k>0$,
the quark potential shows confinement, $E(R)\propto R$ at large $R$. 

Here, using the formula
\eqref{deta} with \eqref{f0} at the zero-temperature case, 
we provide the inverse theorem of it: when the quark potential shows
confinement $E(R) \propto R$ at large $R$,  the bulk needs to have its IR bottom. 

\vspace{5mm}

\begin{screen}
\begin{theorem}
Assume the linear confinement: at large $R$, the quark potential is given by
\begin{align}
\frac{dE(R)}{dR} = \sigma + \frac{c}{R^n} + \mbox{(higher in $1/R$)} \, .
\label{con}
\end{align}
Here $\sigma (>0)$ is the confining string tension.
The second term (with $c>0$ and $n>0$) is the leading correction.

Then the bulk metric function $f(\eta)=g(\eta)$ in \eqref{m} has an IR bottom: $f(\eta)$ approaches
a minimum $f=2\pi\alpha'\sigma$ at which the gradient $df/d\eta$ vanishes.
The location of the IR bottom in the $\eta$ coordinate is 
\begin{itemize}
\item
at a finite value of $\eta$, when $n>2$ 
(or when the correction vanishes faster than the power-law). 
\item 
at $\eta=-\infty$,
when $2\geq n > 0$. 
\end{itemize}
\end{theorem}
\end{screen}

\begin{proof}
%
The confinement condition \eqref{con},\footnote{The positivity of $c$
is natural, since normally the large-$R$ quark potential $E(R)$ approaches its linear behavior
from below, as it is consistent with the short-distance behavior.} 
in which we safely ignore the higher terms
for large $R$,
is substituted to \eqref{f0} to give
\begin{align}
R(f_0) = \left(
\frac{f_0-\tilde\sigma}{\tilde{c}}
\right)^{-\frac1n} \, ,
\label{Rf0}
\end{align}
where $\tilde\sigma\equiv 2\pi\alpha'\sigma$ and $\tilde{c} \equiv 2\pi\alpha' c$.
This is substituted to the right hand side of \eqref{deta}, and is evaluated, 
with the change of the integration variable $f_0 = f x$, as
\begin{align}
\frac{d\eta(f)}{df}
&
= 
\frac{1}{\pi}\sqrt{f}\frac{d}{df}
\int_\infty^1 dx \frac{\tilde{c}^{\frac1n}}{(fx-\tilde\sigma)^{\frac1n}\sqrt{x^2-1}}
\nonumber \\
&
= 
\frac{\tilde{c}^{\frac1n}}{n \pi} f^{-\frac12-\frac1n}
\int_1^\infty dx \frac{x}{\left(x-\frac{\tilde\sigma}{f}\right)^{\frac1n+1}\sqrt{x^2-1}}
\nonumber \\
&
\simeq 
\frac{\tilde{c}^{\frac1n}}{n \pi} f^{-\frac12-\frac1n}
\frac{1}{\left(1-\frac{\tilde\sigma}{f}\right)^{\frac12+\frac1n}}
\frac{\sqrt{\pi/2} \, \Gamma(\frac12+\frac1n)}{\Gamma(1+\frac1n)}
\nonumber \\
&
=\frac{\tilde{c}^{\frac1n}}{n\sqrt{2\pi}}
\frac{\Gamma(\frac12+\frac1n)}{\Gamma(1+\frac1n)} (f-\tilde\sigma)^{-\frac12-\frac1n} \, .
\end{align}
In ``$\simeq$" above, we have used $f \simeq \tilde\sigma$ since $R$ is sufficiently large in \eqref{Rf0}.
The last expression is integrated  to give
\begin{align}
\left\{
\begin{array}{ll}
\displaystyle
\eta = \frac{\sqrt{2}\tilde{c}^{\frac1n}}{(n-2)\sqrt{\pi}}
\frac{\Gamma(\frac12+\frac1n)}{\Gamma(1+\frac1n)} (f-\tilde\sigma)^{\frac12-\frac1n} + \mbox{const.}
& \qquad
(n\neq 2) 
\\[4mm]
\displaystyle
\eta = \frac{\sqrt{\tilde{c}}}{\sqrt{2}\pi}
 \log(f-\tilde\sigma) + \mbox{const.}
& \qquad
(n=2)
\end{array}
\right.
\end{align}
This is easily inverted to give $f(\eta)$, as
\begin{align}
\left\{
\begin{array}{ll}
\displaystyle
f(\eta) = \tilde{\sigma} +  \left(d_n \eta-\mbox{const.}\right)^\frac{2n}{n-2} & \qquad (n\neq 2) 
\\[2mm]
\displaystyle
f(\eta) = \tilde\sigma + \exp\left[\frac{\sqrt{2}\pi}{\sqrt{\tilde{c}}}(\eta - \mbox{const.})\right] 
& \qquad (n=2) 
\end{array}
\right.
\end{align}
where $d_{n> 2}$ ($d_{n<2}$) is a positive (negative) constant. The expression proves the theorem,
and at the IR bottom of the bulk geometry, we find $f=\tilde{\sigma}$.

The theorem is also true for the correction $c/R^n$ replaced by $\exp(- c R^m)$ with $c>0$ and $m>0$,
that is, an
exponentially fast approach to the linear confinement potential. The proof goes in the same manner as above,
and we find
\begin{align}
\eta = e_m \sqrt{f-\tilde{\sigma}} \left(\log \frac{\tilde\sigma}{f-\tilde\sigma}\right)^{\frac1m-1}
+ \mbox{const.}
\end{align}
with a positive constant $e_m$ which depends on $c$ and $m$. Inverting it results in the existence
of the IR bottom $f=\tilde\sigma$ at a finite bulk location in $\eta$.
\end{proof}

The famous holographic geometry showing the confinement is a doubly Wick-rotated black 4-brane solution \cite{Witten:1998zw}.
The IR bottom of the geometry is located at a finite value of $\eta$. In fact, the leading correction
of the holographic Wilson loop using the geometry, evaluated in \cite{Greensite:1998bp}, is exponentially suppressed as $\exp[-cR]$, thus
it is consistent with our theorem.

\section{Discussions}
\label{sec:5}

Various comments and discussions about our formulas and the theorem are in order.

\begin{description}

\item[{\bf L\"uscher term.} \hspace{5mm}]

It would be very interesting to plug in lattice results of Wilson loops to our formulas, to obtain a bulk geometry.
When substituting lattice results, there is a subtlety. The quark potential at large $R$ in confining gauge theories
has a $1/R$ correction called L\"uscher term \cite{Luscher:1980ac}. This correction stems from a quantum fluctuation
of the Nambu-Goto string, and the effect in the gravity dual was evaluated in \cite{Kinar:1999xu}. Note that our formulas
are for a classical string in the bulk, so the L\"uscher term is dropped. Because lattice results include the
L\"uscher term in general (and the L\"uscher term is found to be rather consistently seen in lattice simulations \cite{Luscher:2002qv}), 
in applying the simulation results to our formulas, the L\"uscher term contribution needs to be subtracted in the lattice results. 
To precisely argue the quantum corrections, the relevance to the large $N$ limit and the strong coupling limit may 
need to be worked out.\footnote{In fact, in the holographic setup of pure Yang-Mills theory which is dimensionally reduced 
from 5-dimensional supersymmetric Yang-Mills theory \cite{Witten:1998zw},
the QCD scale and the QCD string tension are parametrically separate by the factor of 't Hooft coupling, so
the subtraction of the L\"uscher term would be a tractable problem. However, in realistic Yang-Mills theory these two scales may
not be distinguishable.}

\item[{\bf Unstable branch of string.} \hspace{5mm}]

In the formula \eqref{h} for the non-zero temperature case, note that the connected fundamental string corresponding to
the temporal Wilson loop with the
minimum energy does not reach the horizon \cite{Brandhuber:1998bs,Rey:1998bq}, therefore the formula \eqref{h} is not
sufficient to build the temporal component of the metric near the horizon.
The phenomena is the Debye screening of color charges, and in the bulk a pair of straight separate strings is favored.
Our formulas work for a single curved string connecting the two boundary points, and 
in general, for a given black hole geometry, 
there exist two such configurations: one is the minimum energy configuration that is favored when $R$ is small.
The other is a local maximum energy configuration that is always unfavored. 
The latter configuration is classical and probes the bulk region
very close to the horizon (see \cite{Hashimoto:2018fkb} for its horizon behavior and the relevance to chaos),
so it could be more appropriate for our purpose.
When one wants to probe $f(\eta)$ of the region very close to the horizon,
one needs this local maximum $E(R)$ which may be hidden in lattice data. This $E(R)$ is related to quantum string breaking and quantum chaos
of Wilson loops, but the details on how to extract the local maximum $E(R)$ from lattice data requires some study.

\item[{\bf Machine learning holography.}\hspace{5mm}]

For solving the inverse problem of building a bulk spacetime from QFT data, recently an effort has been put to utilize machine learning,
where the holographic bulk spacetime is identified with neural networks 
\cite{You:2017guh,Hashimoto:2018ftp,Hashimoto:2018bnb,Hu:2019nea,Hashimoto:2019bih,Tan:2019czc,Yan:2020wcd,Akutagawa:2020yeo,Hashimoto:2020jug}\footnote{
See \cite{Ruehle:2020jrk} for a review of data science approach to string theory, and also see
\cite{Carleo:2019ptp} for applications of machine learning to material sciences.}.
Among the work, the method which uses only QFT spectra or one-point function \cite{Hashimoto:2018ftp,Hashimoto:2018bnb,Tan:2019czc,Yan:2020wcd,Akutagawa:2020yeo,Hashimoto:2020jug} would be suitable for lattice simulation data.
A combination of the machine learning method and the results of the present paper would further constrain possible bulk geometries.

\item[{\bf Distinguishing holographic QFTs from non-holographic QFTs.}\hspace{5mm}]

For a given QFT, it would be ideal if one can judge the existence of a gravity dual by just examining QFT observables obtained in lattice simulations.
A protocol for getting a bulk spacetime geometry from Wilson loops is given in this paper, thus it would be desirable to
calculate holographically some other physical observables with this geometry and compare them with lattice data of the same QFT.
For example, non-rectangular Wilson loops (such as circular Wilson loops) would be the first choice, as the holographic computation of 
it needs only the string frame metric in the bulk. If the holographic result is different from the lattice result,
then it means that there is no consistent bulk gravity dual. 
This kind of falsification test of the existence of the gravity dual will help developing study
of possible borders between holographic QFTs and non-holographic QFTs.

\end{description}

%


\let\doi\relax

\end{document}